# A quasi-linear spin-torque nano-oscillator via enhanced negative feedback of power fluctuations

O. J. Lee,[1] P. M. Braganca,[1] V. S. Pribiag,[1] D. C. Ralph[1,2] and R. A. Buhrman[1*]

[1]Cornell University, Ithaca, NY, 14853, USA
[2]Kavli Institute at Cornell, Ithaca, NY, 14853, USA

We report an approach to improving the performance of spin torque nano-oscillators (STNOs) that utilizes *power-dependent negative feedback* to achieve a significantly enhanced dynamic damping. In combination with a sufficiently slow variation of frequency with power this can result in a *quasi-linear* STNO, with very weak non-linear coupling of power and phase fluctuations over a range of bias current and field. An implementation of this approach that utilizes a non-uniform spin-torque demonstrates that highly coherent room temperature STNOs can be achieved while retaining a significant tunability.

[*] buhrman@cornell.edu

In a spin-torque nano-oscillator (STNO) a spin-polarized current (*I*) excites persistent magnetic precession at microwave frequencies in an unpinned magnetic element, the free layer (FL), when the anti-damping spin torque ($\tau_{st}$) is sufficient to compensate the magnetic damping torque ($\tau_d$) [1,2,3,4,5,6]. A seemingly attractive feature of STNOs is their high agility, *i.e.* a strong variation of oscillation frequency *f* with oscillator power, but this, as pointed out by the non-linear auto-oscillator (NLAO) analysis [2,7,8,9], couples thermally-generated amplitude and phase fluctuations, which degrades phase stability (broadens the oscillator linewidth *Δf*).

Here we present a method for achieving enhanced phase stability in a STNO device which utilizes a magnetic configuration that naturally implements enhanced *negative feedback* of oscillator power fluctuations and hence achieves a high effective dynamic damping. When employed in a STNO design that under appropriate bias conditions also exhibits single mode behavior and a relatively low agility, this results in a low field, quasi-linear STNO with a room temperature *Δf* ≈ 5 MHz, very close to that predicted for a linear STNO with the same oscillator energy.

Several conditions are necessary for achieving narrow STNO linewidths. First, to eliminate mode jumping and other mode-mode interactions that invariably broaden linewidths [10,11], the STNO must exhibit single mode excitation. Second, the phase stability of the mode should be maximized. Initial guidance for this is provided by the NLAO analysis [12,13,14,15] which concludes that the amount by which amplitude fluctuations of the STNO renormalizes its thermally-generated intrinsic phase noise is determined by a nonlinear coupling factor (*v*). This leads to the prediction [2,7,8] that Δ*f* is, in the regime where $\Delta f \ll p_0 \Gamma_{eff} / 2\pi$ (quantities defined below),

$$\Delta f = (\Gamma_G / 2\pi)[k_B T / E_0](1+v^2) = \Delta f_0(1+v^2),  \quad (1)$$

Here $\Gamma_G \approx \Gamma_0 \equiv \alpha_G \gamma 2\pi M_S$ (for low field in-plane precession) where $M_s$ is the saturation magnetization of the oscillating magnetic FL, $\alpha_G$ is the Gilbert damping parameter, $k_B$ is the Boltzmann constant, $T$ is the temperature, and $\gamma$ is the electron gyromagnetic ratio, and $E_0 = \beta p_0$ is the time averaged magnetic energy of the oscillation, where $\beta$ depends on the mode of oscillation (see supplementary material [16], S.1), and $p_0$ ($\leq 1$) is the normalized steady-state power of the magnetic oscillation ($p_0 \approx \sin^2(\varepsilon/2)$ where ε is the in-plane maximum excursion angle from the precession axis). The coupling factor $v$ is defined as

$$v \equiv N / \Gamma_{eff} \equiv 2\pi \, df / dp_0 / \left[ \partial \Gamma(p,I) / \partial p \right]\big|_{p_0}, \quad (2)$$

where $p$ is the instantaneous normalized power (which may fluctuate away from $p_0$), $N \equiv 2\pi \, df / dp_0$ is the agility of the STNO (assumed to be independent of $I$, except through $p_0(I)$), and the total damping $\Gamma(p,I) = \Gamma_+(p) - \Gamma_-(p,I)$ is the difference between the natural positive dissipative damping, $\Gamma_+(p) = \Gamma_0(1 + \eta(p_0))$, and the net anti-damping caused by $\tau_{ST}$, $\Gamma_-(p,I) = \sigma_0 g(p) I$. Here $\eta(p_0)$ represents any nonlinear behavior of the dissipative damping, which a recent experiment [11] has indicated is negligible at moderate $p_0$ so we will assume hereafter $\eta(p_0) \approx 0$, and $g(p)$ describes the power dependence of the anti-damping (with $g(0)=1$) and $\sigma_0 = \Gamma_0 / I_c$, with $I_c$ the critical current required for the onset of oscillation.

This NLAO analysis indicates that $\Delta f$ is minimized when both $E_0$ is maximized and $|v|$ is minimized, with ideally $|v| \leq 1$, which leads to strategies [17,18,19] for the reduction of $|v|$ through the application of either an in-plane hard axis or an out-of-plane magnetic field, to bring $N$ close to zero by balancing opposing contributions of different anisotropy fields. Experiments [20,21,22,23,24,25,26] have demonstrated that a smaller $\Delta f$ can indeed be obtained in this way although in most cases $df/dI$ rather than $N$ is reported. Measurement of $df/dI$ yields the STNO's non-linear coupling constant since we also have ([16], S.2)

$$\nu = (2\pi/\Gamma_0) I (df/dI) , \qquad (3)$$

but does not reveal whether a lower $\nu$ is due to a low $N$ or to an unusually high $\Gamma_{eff}$. $\Gamma_{eff}$ can be determined from the variation of $p_0$ with bias current ([16], S.2),

$$\Gamma_{eff} = \Gamma_0 \left( I \, dp_0/dI \right)^{-1}, \qquad (4)$$

but such determinations are usually not reported. Typically $p_0 \approx (I - I_c)/I$, for $I_c \leq I < 2I_c$. In such a case, Eq. (4) indicates that $\Gamma_{eff} \leq 2\Gamma_0$, and this has been directly found to be the situation in a quantitative study [11] of a spin valve STNO where the precession axis of the FL was collinear with the reference layer (RL). Thus, given typical values of $\Gamma_0/2\pi \approx 0.1$ GHz, to achieve $|\nu| \leq 1$ the agility has to be tuned also to the low value $|N|/2\pi < 0.2$ GHz, which macrospin modeling indicates can at best be accomplished over only a very narrow field bias range ([16] S.3). Nonetheless there are recent reports [15,25] of STNOs with $|\nu| \leq 3$. Below we show that a quasi-linear ($|\nu| \leq 1$) STNO can be realized by engineering a strong negative feedback for oscillator power fluctuations to greatly increase $\Gamma_{eff}$, while maintaining a small value of $|N|$ and single mode oscillation.

To demonstrate this strategy we fabricated ([16] S.4) tapered nanopillar spin-valve STNOs with the layer structure Py(5)/Cu(12)/Py(20) (thicknesses in nm, Py = $Ni_{80}Fe_{20}$), and with the thin (5 nm) FL located closer to the substrate than the thicker ferromagnetic RL [27,28]. The devices are patterned by electron-beam lithography and Ar ion milling to have a FL cross-section approximately 50 × 145 nm$^2$, with sidewalls tapered ~ 20-30° from vertical. The dipolar field ($H_d$ ≈ 250 Oe) from the RL acts to orient the FL anti-parallel (AP) to it, giving an offset angle, as averaged over the FL, $\langle \varphi_0 \rangle$ = 180° in zero field. We apply an in-plane hard axis magnetic field, $H_y$ ≤ 1000 Oe, that acts to reduce $\langle \varphi_0 \rangle$ near to but not past 90°. Because we still have $\langle \varphi_0 \rangle$ > 90°, FL oscillations are generated for $I < 0$ such that electrons flow from the RL to the FL. Here we report room temperature (RT) results from one particular device but the behavior was quite similar for all 4 devices that were measured in detail, with differences attributed to device geometry variations in the fabrication process.

Figure 1a shows the measured power spectral densities (PSDs) of a device at $H_y$ = 520, 610, 700, 820 and 880 Oe for $I$ = -4 mA. For values of $H_y$ smaller than 650 Oe the STNO exhibited two or more modes with broad linewidths due to mode jumping (for the 610 Oe curve there is a second small peak near $f$ = 4.8 GHz), which we attribute to the relatively wide spatial distribution of the natural oscillation frequency associated with variations in the internal field across the FL (see ref. [28]). At and above $H_y$ = 650 Oe, a single mode is observed, with $f$ increasing with $H_y$ as expected for in-plane oscillation. In this regime the primary signal power is at the fundamental precession frequency.

In Fig. 1b-d we show, for the single mode regime with $H_y$ = 700, 820, and 880 Oe, the values of $f$, $p_N$ (= $P / I^2$ yielding the power of the underlying resistance oscillations, where $P$ is the measured STNO output microwave power), and $\Delta f$ as a function of $|I|$, as obtained from Lorentzian fits to the measured PSDs. The key finding is that $\Delta f$ goes through broad minima as a function of |I| (see the $H_y$ = 700 Oe curve in Fig. 1d) with the lowest RT linewidths $\Delta f \approx$ 5 MHz. Unlike most previous experiments in which narrow linewidths were observed only within very narrow field ranges, we observe $\Delta f \leq$ 8 MHz for $H_y$ ranging from 650 to 800 Oe, with the minima shifted to larger values of |I| for the larger $H_y$ (Fig. 1d; 5 mA was the largest value of current bias used in the experiment). Fig. 1b shows that for $H_y$ between 700 and 880 Oe, the oscillation $f$ undergoes a crossover from a red shift with |I| ($H_y$ = 700 Oe) to a blue shift ($H_y$ = 880 Oe) (Fig. 1b), so that $df/d|I|$ passes through zero in this field bias range. Thus as expected, the bias region where low values of $\Delta f$ are obtained is associated with relatively small values of $df/d|I|$ and therefore $|v|$, with the best results depending on the optimum combination of a low $df/d|I|$ and a high $p_N$.

Fig. 1c shows that at sufficiently high |I| $dp_N/d|I| \to 0$, which is an indication by Eq. (4) that $\Gamma_{eff}$ is diverging. When $H_y$ is changed, the onset current for the beginning of the saturation in $p_N$ (Fig. 1c) shifts in the same fashion as the onset current for the range of minimum $\Delta f$ (Fig. 1d). The value of $p_N$ at saturation also depends on $H_y$; as $H_y$ moves the initial offset angle between the FL and the RL closer to 90°, $p_N$ at saturation decreases.

A quantitative comparison with the NLAO model requires scaling $p_N$ to estimate the normalized oscillator power $p_0$, which we do by employing the macrospin approximation as discussed in Ref [16], S.5. In Fig. 2a we plot $\Delta f \times p_o$ as determined in this manner from the measured data of Fig. 1b-d. Over a wide range of $I$ and $H_y$ we see that while $p_o$ varies by a factor of 30, $\Delta f \times p_0$ varies by $\leq$ 2 and is within a factor of two of the predicted linear oscillator value

$\Delta f_{\text{pred}} \times p_o \approx 0.19$ MHz (for $v = 0$, $f = 6$ GHz, $\alpha_G = 0.01$, $M_S = 560$ emu/cm$^2$, $V = 2.9 \times 10^4$ nm$^3$), in accord with Eq. (1).

Figures 2b and 2c show, respectively, $p_o$ and $\Delta f_{meas}$ as a function of $H_y$ for $I = -4$ mA. As $H_y$ increases from 600 Oe $\Delta f_{meas}$ decreases rapidly, reaching its minimum $\approx 5$ MHz at $H_y^{opt} \approx 700$ Oe, and then increases again with higher $H_y$, while $p_o$ decreases monotonically. Also shown in Fig. 2c is the NLAO prediction $\Delta f_{pred}$, as determined from Eqs. (1) and (3) using the estimated values of $p_o$ and the measured values of $I\, df/dI$ ([16] S.5). There is quite good agreement between experiment and the prediction for $H_y$ from 675 Oe to 800 Oe. [The deviation of $\Delta f_{pred}$ above $\Delta f_{meas}$ for $H_y$ above 800 Oe is attributable to the low power restoration rate [7,11] $\Gamma_p\ (= p\Gamma_{eff})$ there ([16] S.6)]. In Fig. 2d we plot $\Gamma_{eff}/2\pi$ vs. $I$ for $H_y = 700$ Oe. Below $I_c$ ($\approx$ -2.6 mA) $\Gamma_{eff}$ decreases rapidly with increasing $I$, plateaus for $I \approx I_c$ at $\Gamma/2\pi \sim 1$ GHz, and then increases again as $p_o$ begins to saturate with $I$.

We can utilize the NLAO model to determine $v$ using Eq. (3), obtaining the results shown in Fig. 2e. We achieve quasi-linear behavior ($|v| \leq 1$) over a significant range of $H_y$ and $I$. We determine the individual contributions of $N$ and $\Gamma_{eff}$ to this result by evaluating $N/2\pi = df/dp_0$ using the measured values of $f$ and the estimated values of $p_0$ and by evaluating $\Gamma_{eff}/2\pi$ from Eq. (4) (Fig 2f). While the values of $N/2\pi$ that we find in the optimal regime, where there is a gradual red shift to blue shift transition in the agility as a function of $H_y$ (see [16] S.3), are in a range comparable to a previous STNO experiment of a different design [11] ($N/2\pi \sim 1$ GHz), the values of $\Gamma_{eff}$ in our experiment ($\Gamma_{eff}/2\pi \sim 1$-3 GHz), are more than ten times greater.

We have performed micromagnetic simulations (MMS) (see [16] S.7) to gain insight into the origin of the enhanced dynamic damping. First, to confirm the applicability of the MMS results to our device, in Fig. 3a we show the power of the MMS oscillation $p_{o,mms}$ as a function of $I$ for $T = 0$ and $H_y = 800$ Oe, which exhibits both a gradual onset above $I_c$, and the quasi-saturation behavior seen experimentally (Fig. 1c.) (We attribute the quantitative differences in behavior to differences in the actual size and shape of the device, and the measurements were performed at room temperature, not $T = 0$.)

In Fig. 3b we show the static solution for $H_y = 800$ Oe and $I = 0$ which indicates a significant out-of-plane magnetization (OPM) component in the bottom end regions of the RL, up to 12%. The MMS suggests that this OPM plays a significant role in the dynamics because the current incident on the FL end regions has a significant out-of-plane polarization (OPP) component that acts to induce the magnetization of the right (left) FL end region to have a significant time-averaged +z (-z) component. For $I \simeq I_c$ the principal effect is that the in-plane (IP) "clam-shell" precession [3] of the FL does not proceed uniformly as it would for a rigid domain with a purely IP ST. Instead the $M_z = 0$ crossings of the IP oscillation are non-uniform in time across the FL (see Fig. 3c and [16], Movie S.1 for $I = -2.25$ mA), due to the opposite "pinning" of the time-averaged OPM in the two ends, $\langle M_{z,L} \rangle < 0$ and $\langle M_{z,R} \rangle > 0$. The MMS indicates that this phase difference reduces the ST effectiveness for a given $I$ (reduces the oscillation energy) in comparison to a fully in-phase case, and thus retards the growth of $p$ (oscillation amplitude $\varepsilon$) with $I$ just above $I_c$ (enhances $\Gamma_{eff}$.)

As $\varepsilon$ increases further with $I$ the stronger exchange coupling of the oscillation across FL reduces the fraction of the cycle where $\text{sgn}(M_{z,L}) \neq \text{sgn}(M_{z,R})$ and $\Gamma_{eff}$ is reduced somewhat (see

Fig. 3a and 3d, and [16], Movie S.2 and S.3, for $I$ = -2.5 mA and – 3 mA). Eventually however, for sufficiently high $I$, $\varepsilon$ in the end regions becomes large enough that the extremum of the IP offset angle $\varphi_{min} = \varphi_0 - \varepsilon$ is reduced to ~ 90° relative to the RL. At that point the instantaneous IP ST exerted on the FL has its maximum strength. When $I$ is increased further there is a decreasing amount of net "anti-damping" torque on the FL for increasing precession amplitude, while $\tau_d$ continues to grow with $\varepsilon$. Moreover, the MMS indicates that once $\varphi$ varies to below 90° during the oscillation the "back-action" ST exerted on the RL by the anti-parallel spin-polarized electrons reflected from the FL is sufficient, due to the low demagnetization field in the end regions of the RL, to set those regions into significant precession about their local effective fields, thus periodically increasing and decreasing the out-of-plane polarization of the current incident on the FL. The MMS indicates that this makes the large-amplitude in-plane FL precession less uniform across the FL, which strongly reduces the efficiency of the IP ST (See Fig. 3e and [16], Movie S.4). The effect is a strong enhancement of $\Gamma_{eff}$ for $|I| > 3$ mA in the MMS, while the time and spatially averaged MMS oscillation remains relatively coherent.

We suggest that the enhanced damping illustrated in the MMS and observed experimentally is a natural consequence of the non-collinear and non-uniform magnetic configuration in our device, and that the OP ST component that arises from the bottom of the tapered ends of the RL is principally responsible for the negative feedback (*i.e.*, large $\Gamma_{eff}$) that suppresses power fluctuations, an effect that cannot be captured in macrospin modeling. When the bias is increased sufficiently above threshold that $\varphi$ becomes $\leq 90°$ in the end regions during the extremum of the cycle, the magnitude of the ST interactions between the FL and the less stable end regions of the RL is enhanced, resulting in momentarily stronger OPM there. This leads to a further increase in the strength of the negative feedback – a fluctuation to a larger instantaneous precessional power ($\varepsilon$) results in a rapid suppression of the fluctuation. Mathematically, for this device the anti-damping caused by the IP ST, $\Gamma_-(p,I)$, becomes a strongly decreasing function of $p$ near values of $p$ corresponding to $\varphi_{\min} \approx 90°$, leading to the large enhancement of $\Gamma_{\textit{eff}} \approx -\partial \Gamma_-(p,I)/\partial p$. This differs from the case of collinear STNOs with uniform ST where the anti-damping is predicted to have the simple form [2,11]

$\Gamma_-(p,I) = \sigma_0(1-p)I$.

In summary we have shown how an enhanced dynamic damping can reduce the nonlinear coupling that hinders the phase stability of conventional STNOs, and have also demonstrated that an enhanced $\Gamma_{\textit{eff}}$ can be realized by employing a tapered, non-collinear device configuration, where an OP ST component can provide a strong negative feedback control that restricts power fluctuations. An attractive feature of this STNO is that while $|\nu| \leq 1$, it is still frequency tunable (Fig. 1b) to a useful degree, *i.e.* with $|\Delta I \, df/dI| \gg \Delta f$, due to the change of $\langle \varphi_0 \rangle$ with $I$ (Fig. 3a).

**Acknowledgements**

This work was supported in part by ONR, DARPA, ARO, and NSF/NSEC through the Cornell Center for Nanoscale Systems. We also acknowledge support from the NSF through use of the Cornell Nanofabrication Facility/NNIN and by NSF/MRSEC (DMR-1120296) through use of the Cornell Center for Materials Research (CCMR) facilities.

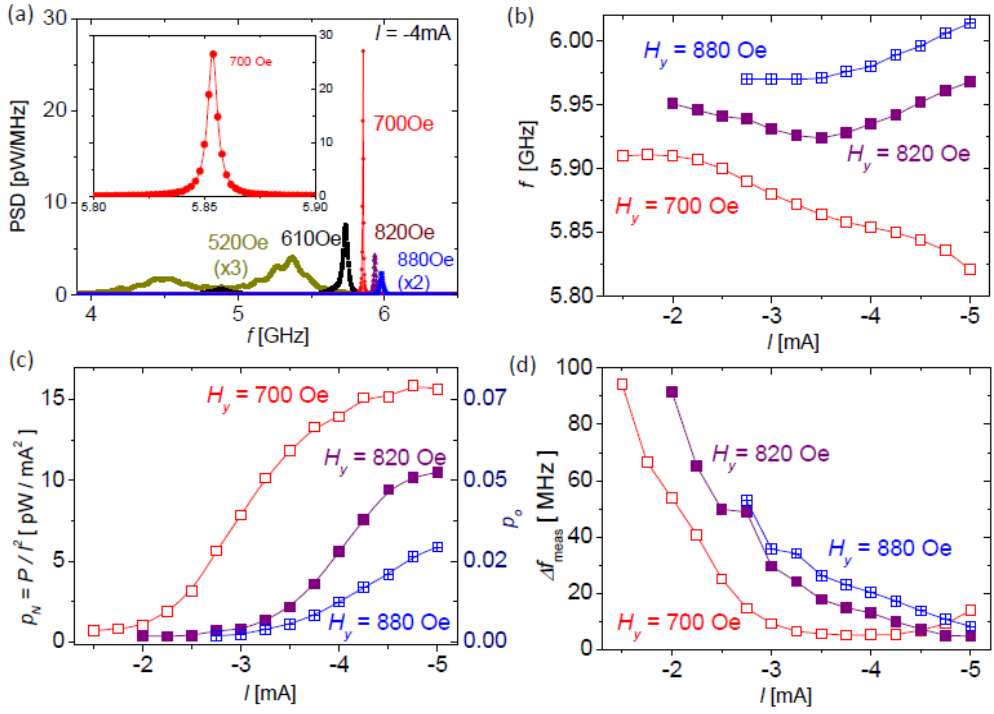

**Fig. 1** (a) Measured PSDs for $H_y$ = 520, 610, 700, 820 and 880 Oe at $I$ = -4 mA; (b) oscillator frequency $f$; (c) oscillator power $p_N$ ($=P/I^2$) and normalized power $p_o$; and (d) $\Delta f_{meas}$ as a function of $I$, all for $H_y$ = 700 Oe, 820 Oe and 880 Oe.

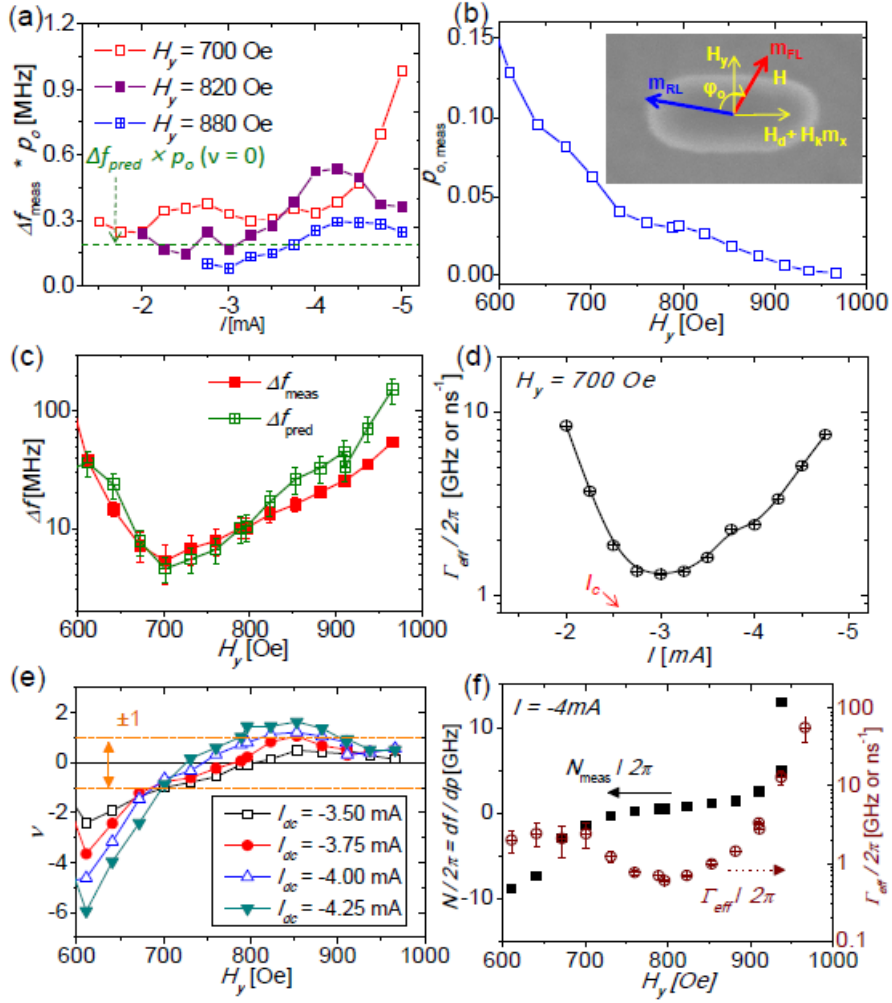

**Fig. 2** (a) Measured $\Delta f_{meas} \times p_0$. (b) Normalized oscillator power $p_0$ as a function of $H_y$ for $I$ = -4 mA. Inset: Schematic of the internal/external fields and the magnetic configuration of the STNO. (c) Measured oscillator linewidth $\Delta f_{meas}$ for $I$ = -4 mA. Also shown is the predicted linewidth $\Delta f_{pred}$ from Eqs. (1) and (3) using the values of $p_0$ shown in (b) and the measured values of $I\, df/dI$. (d) Effective damping vs. $I$ for $H_y$ = 700 Oe. Up to $I_c$ ($\approx$ -2.6 mA) $\Gamma_{eff}/2\pi$ drops rapidly with increasing $I$ due to the anti-damping ST effect on the thermally generated oscillation. Above -3.25 mA, $\Gamma_{eff}/2\pi$ increases approximately linearly as the variation of oscillation power with bias becomes weaker and weaker. (e) Non-linear coupling constant $\nu$ as determined from $I\,df/dI$ and Eq. (3) vs. $H_y$ at several different bias currents $I$.

(f) $N/2\pi = df/dp_0$ and $\Gamma_{eff}/2\pi$ of the STNO vs $H_y$ for $I = -4$ mA.

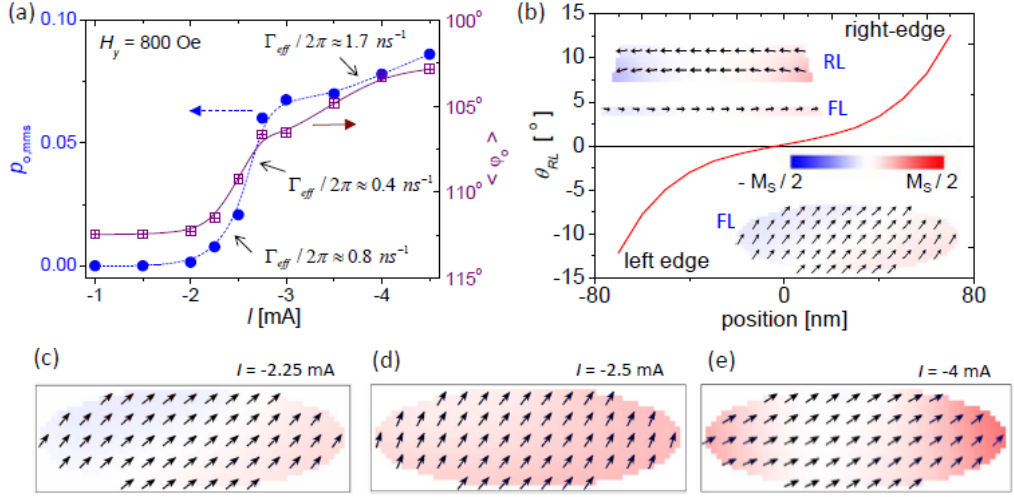

**Fig. 3** (a) Normalized oscillator power $p_{o,mms}$ vs. $I$, as determined from the MMS for $H_y = 800$ Oe. Also shown is the mean offset angle $\langle \varphi_0 \rangle$ averaged over the simulated FL. (b) The out-of-plane orientation ($\theta_{RL}$) of the RL magnetization as the function of position along its elongated (easy) axis, as determined by MMS for $H_y = 800$ Oe. **Insets:** Cross-section view of the static ($I = 0$) magnetization configuration of the tapered STNO and top view of the FL. The color scaling indicates the local out-of-plane $M_z$ component. (c) MMS snapshot of the FL magnetization at the approximate mid-point of a precession for $I = -2.25$ mA. The in-plane precession is in the direction of increasing $\varphi$. (d) MMS FL snapshot at the approximate mid-point of a precession for $I = -2.5$ mA. (e) MMS FL snapshot at the approximate mid-point of a precession for $I = -4.0$ mA.

# A quasi-linear spin-torque nano-oscillator via enhanced negative feedback of power fluctuations

O. J. Lee,[1] P. M. Braganca,[1] V. S. Pribiag,[1] D. C. Ralph[1,2] and R. A. Buhrman[1]

[1]Cornell University, Ithaca, NY, 14853, USA
[2]Kavli Institute at Cornell, Ithaca, NY, 14853, USA

## SUPPLEMENTARY INFORMATION

**Contents:**

S.1. STNO magnetic energy and positive damping rate for the case of in-plane precession

S.2. Derivation concerning the nonlinear effective damping ($\Gamma_{eff}$) of a STNO

S.3. Macrospin estimates of the STNO agility $N/2\pi \equiv df/dp$ and of oscillator power $p_0$ for different field bias conditions

S.4. Tapered spin valve STNO fabrication process

S.5. Experimental estimation of the STNO's normalized oscillator power $p_0$ and nonlinear coupling $v$

S.6. Bias dependence of the power restoration rate $\Gamma_p$

S.7. Micromagnetic simulations of the STNO

**S.1. STNO magnetic energy and positive damping rate for the case of in-plane precession**

The time-averaged magnetic energy of a STNO can be written as $E_0 = \beta p_0$ where $p_0 = \sin^2(\varepsilon/2)$ is the normalized steady-state power, $\varepsilon$ is the maximum in-plane precession angle, and the factor $\beta$ depends upon the mode of precession. In this section we determine $\beta$ for the geometry relevant to our experiment (see inset in Fig. 2b of the main text). A version of this calculation has been performed previously [1] for the case of out-of-plane circular precession that occurs in the presence of a strong out-of-plane applied field $H_z$. In that case one has $\beta = 2\pi M_s V f / \gamma$ where the oscillator frequency $f$ varies linearly with $H_z$.

However, in our experiment the free layer precesses about an axis that is in the sample plane. For this case, the orbit of the FL moment is elliptical due to the effect of the strong out-of-plane demagnetization field, and $f$ does not have a linear dependence on the applied field. The magnetic energy per unit volume, $u_M (\equiv E(\varphi,\theta)/V)$, of the FL, under the assumptions that the applied field $H_y$ is oriented along the in-plane hard axis and that the RL does not rotate at all with $H_y$, is given by

$$u_M = -H_d M_s \cos\theta\cos\varphi - H_y M_s \cos\theta\sin\varphi - (H_k M_s/2)\cos^2\theta\cos^2\varphi + 2\pi M_s^2 \sin^2\theta, \quad (S1)$$

where $\theta$ is the out-of-plane tilt angle of the FL moment $\vec{m}$ and $H_k$ is the in-plane anisotropy field. For a given $H_y$, the equilibrium FL direction lies in the plane at an angle $\varphi_0$ from the RL, and we can define an effective field $\vec{H}$ along the equilibrium axis such that

$$H\sin\varphi_o = H_y \quad \text{and} \quad H\cos\varphi_o = H_k \cos\varphi_o - H_d. \quad (S2)$$

The energy density determines the small amplitude ferromagnetic resonance frequency $f$ of the FL via (see Appendix I in ref. 1 for details)

$$f^2 = \left(\frac{\gamma}{2\pi M_s \cos\theta}\right)^2 \left[\frac{\partial^2 u_M}{\partial \varphi^2}\frac{\partial^2 u_M}{\partial \theta^2} - \left(\frac{\partial^2 u_M}{\partial \varphi \partial \theta}\right)^2\right]_{\theta=0}. \quad (S3)$$

Employing Eq. (S1) we obtain

$$f^2 \approx \left(\frac{\gamma}{2\pi}\right)^2 \left[H - H_k \sin^2\varphi_o\right]\left[H + 4\pi M_s\right]. \quad (S4)$$

Thus $f$ is determined by the effective in-plane anisotropy field ($H_{k,eff} = -H_k \sin^2\varphi_0$) and by the out-of-plane demagnetization field ($4\pi M_S$). The oscillator energy corresponding to a maximum in-plane excursion angle ε (such that $\varphi = \varphi_o + \varepsilon$) is $E_0(\varepsilon) = E(\varphi_o + \varepsilon) - E(\varphi_o)$, assuming that the precession axis does not shift substantially with oscillation amplitude $\varepsilon$ (correct in the low power limit). After some algebra, we find

$$E_0(\varepsilon) \approx \frac{1}{2}[H - H_k \sin^2\varphi_o]M_s V \varepsilon^2. \quad (S5)$$

Therefore, in the low power limit for the experimentally relevant case of precession about an in-plane axis with $H \ll 4\pi M_s$, we have the final result

$$\beta \equiv \frac{E_0(\varepsilon)}{p_0} = \frac{E_0(\varepsilon)}{\sin^2(\varepsilon/2)} \approx \frac{2\pi V f^2}{\gamma^2}. \quad (S6)$$

With respect to the positive magnetic damping rate $\Gamma_G$, it also depends of course on the mode of precession and hence the magnetic field configuration [1,6]. It has been shown [6] however that for in-plane precession and, as in the case of study here, when the applied in-plane fields, $H_y$, $H_k$ and $H_d$, are much less than the out-of-plane demagnetization field ($\approx 4\pi M_s$) the variation of $\Gamma_G$ with bias is quite weak and that $\Gamma_G \approx \Gamma_0 \equiv \alpha_G \gamma 2\pi M_S$. Following the macrospin methodology of Ref. 6 we have calculated $\Gamma_G$ and indeed find it varies by < 20 % from $\Gamma_0$ over our bias range. Hence for simplicity we assumed $\Gamma_G = \Gamma_0$ in this work.

## S.2. Derivation concerning the nonlinear effective damping ($\Gamma_{eff}$) of a STNO

Starting from the Landau-Lifshitz-Gilbert Slonczewski (LLGS) equation [2], the non-nonlinear auto-oscillation model [1,3] derives the following governing equation for a single mode STNO:

$$\frac{\partial a}{\partial t} + i\omega(p)a + [\Gamma_+(p) - \Gamma_-(I,p)]a = f_n(t). \tag{S7}$$

Here $\omega(p) = \omega_0(1+\xi(p))$ is the power-dependent oscillation frequency, $\Gamma_+(p) = \Gamma_o(1+\eta(p))$ is the dissipative positive damping, and $\Gamma_-(I,p)$ is the anti-damping from the spin-transfer-torque, which can generally be expressed as $\Gamma_-(I,p) = \sigma_o I g(p)$ where $g(p=0) = 1$ at $I = I_c$ (and $I_c = \Gamma_o / \sigma_o$). The fluctuating forcing term $f_n(t)$ allows for the inclusion of thermal fluctuation effects. The total time-averaged damping is

$$\Gamma(I,p) \equiv \Gamma_+(p) - \Gamma_-(I,p) = \Gamma_o(1+\eta(p)) - \sigma_o I g(p) = \Gamma_o\left[1 + \eta(p) - g(p)I/I_c\right]. \tag{S8}$$

For a *stationary* (steady-state) auto-oscillation, having an average oscillator power equal to $p_0(I)$, the total time-averaged damping is necessarily zero:

$$0 = \Gamma_o\left[1 + \eta(p_0) - g(p_0)I/I_c\right]. \tag{S9}$$

From Eq. (S9) we have an implicit expression that can be used to determine the steady-state power $p_0$ at a given value of $I$,

$$g(p_0) = \left(1 + \eta(p_0)\right)\frac{I_c}{I}. \tag{S10}$$

Taking a partial derivative of Eq. (S8), we also have

$$\frac{\partial \Gamma(I,p)}{\partial I} = -\Gamma_o \frac{g(p)}{I_c} \, . \tag{S11}$$

Now we wish to determine the effective dynamic damping $\Gamma_{eff} = \partial \Gamma(I,p)/\partial p$, evaluated at $I$ and $p_0$. For points along the trajectory where $p = p_0(I)$, $\Gamma(I,p)$ is identically zero, so that on this trajectory $d\Gamma(I,p) = \frac{\partial \Gamma}{\partial I} dI + \frac{\partial \Gamma}{\partial p} dp_0 = 0$ and therefore

$$\Gamma_{eff} \equiv \frac{\partial \Gamma(I,p_0)}{\partial p} = -\frac{\partial \Gamma}{\partial I} \cdot \left( \frac{dp_0(I)}{dI} \right)^{-1} . \tag{S12}$$

Using Eq. (S11) and (S12),

$$\Gamma_{eff} = \frac{\Gamma_o}{I_c} g(p_0) \left( \frac{dp_0(I)}{dI} \right)^{-1} = \Gamma_o \frac{1+\eta(p_0)}{I} \left( \frac{dp_0(I)}{dI} \right)^{-1} . \tag{S13}$$

so that $\quad v \equiv \dfrac{N}{\Gamma_{eff}} = \dfrac{2\pi \dfrac{df}{dp_0}}{\Gamma_o \dfrac{1+\eta(p_0)}{I}\left(\dfrac{dp_0}{dI}\right)^{-1}} = \dfrac{2\pi}{\Gamma_o} \dfrac{I}{1+\eta(p_0)} \dfrac{df}{dI} . \tag{S14}$

Recent work [4] has found that $\eta(p)$ is negligible in a collinear spin valve STNO, and given the low normalized power levels in the single mode regime in our measurements we assume this also to be the case for our STNO, resulting in:

$$\Gamma_{eff} = \Gamma_o \left( I \frac{dp_0}{dI} \right)^{-1} \quad \text{and} \quad v = \frac{2\pi}{\Gamma_o} \left( I \frac{df}{dI} \right) . \tag{S15}$$

**S.3. Macrospin estimates of the STNO agility** $N/2\pi \equiv df/dp$ **and of oscillator power** $p_0$ **for different field bias conditions**

To obtain a quasi-linear STNO it is necessary that the magnitude of its agility should be no larger than comparable to the effective damping of the oscillator $\Gamma_{eff}$; in our case this requires $N/2\pi = df/dp \leq 2$ GHz. This is achieved by biasing the oscillator with some combination of external and internal effective fields such that the red and blue frequency shifts as a function of oscillator amplitude that arise from different anisotropy field effects are approximately balanced. For an in-plane magnetized spin valve or magnetic tunnel junction STNO analytical approximations [5,6] that treat the FL as a rigid domain have proposed that low agility can be achieved by applying an external field with a substantial hard axis component so that the red shift due to the power dependence of the oscillation frequency from the out-of-plane demagnetization field is balanced by the blue shift arising from the in-plane anisotropy field.

As a check on the predictions of such analytical analyzes we have employed macrospin simulations to estimate $N/2\pi$ utilizing the LLGS equation

$$\frac{d\hat{m}}{dt} = -\gamma \hat{m} \times \vec{H}_{eff} + \alpha_G \hat{m} \times \frac{d\hat{m}}{dt} + \gamma a_J(\varphi)\, \hat{m} \times \hat{p} \times \hat{m} \qquad (S16)$$

where $a_J(\varphi) = \frac{\hbar}{2e}\frac{I}{M_s V} P \frac{g(\varphi)}{2}$ ($g(\varphi)$ represents any orientation dependence of the spin torque due to spin accumulation effects) and $\vec{H}_{eff} = \left(H_k m_x + H_d\right)\hat{x} + H_y \hat{y} - 4\pi M_s m_z \hat{z}$. Here $\gamma$ is the gyromagnetic ratio, $\hat{m} = (m_x, m_y, m_z)$ is the unit vector of the FL, $H_k$ is the anisotropy field of the FL along the easy axis, $H_d$ is the dipole field from the RL, and $H_y$ is the externally applied hard axis magnetic field. In the macrospin simulation we assumed for simplicity that $g(\varphi) = 1$ and the spin-polarization is $P = 0.37$. With Eq. (S16) we obtained $\hat{m}(t)$ and from this time dependence of the rigid FL moment we determined its precessional axis ($\varphi_{FL}$ and $\theta_{FL}$) and $p = \sin^2(\varepsilon/2)$ as a function of $I$ and $H_y$ for $H_d = 250$ Oe and $H_k = 450$ Oe. Here $\varphi_{FL}$ is the in-plane orientation of the FL as calculated from $-\hat{x}$, $\theta_{FL}$ is its out-of-plane tilt angle, and $\varepsilon$ is the maximum in-plane precessional excursion angle (see Fig. S1b). (A nonzero value of $\theta_{FL}$ results in our non-collinear device geometry from the non-zero time-averaged spin torque oriented in the sample plane and perpendicular to the precession axis of the free layer – the free layer tilts slightly out of plane so that in equilibrium the torque from the demagnetization field can cancel this spin torque to give zero net torque.)

The macrospin simulation result for $N/2\pi$ at the onset of oscillation for the case of $H_k = 450$ Oe, $H_d = 250$ Oe is shown in Fig. S1a, along with the agility as predicted by a small-amplitude oscillation analytical approximation [6] to the LLGS equation. It is initially a bit surprising that analytical approximation agrees reasonably well with the full macrospin modeling, albeit only in the high $H_y$ regime, since the former assumes a circular orbit while an elliptical orbit is expected for ST driven precession in the small amplitude limit, at least from macrospin simulations in the collinear case [7]. Our simulations indicate that this is due to the fact that in the strongly non-collinear regime, the non-conservative, $\varphi$-dependent ST affects the time-averaged dynamic energy balance of the precessional orbit and significantly reduces its ellipticity, as illustrated by Fig. S1b. However when we compare the macrospin predictions with the measured results that are also shown in Fig. S1a for the STNO discussed in the main text we see that there is not good quantitative agreement with the experimental agility, either as measured right above the onset of oscillations, or as measured at a fixed bias current $I = -4$ mA where experimentally $|v| \leq 1$ for $H_y \geq 700$ Oe. While the macrospin modeling does provide a qualitative understanding of the red to blue shift transition in agility as the function of hard axis bias it does not provide precise guidance as to the combination of bias and anisotropy fields that balances $N$ close to zero in our STNOs.

Our macrospin modeling also provides some insight into how different orientations of the FL magnetization relative to the incident spin polarization direction $\hat{p}$, which is determined by the local RL magnetization, affect the local ST efficiency. In general, $\vec{\tau}_{st}(t)$ is in the direction $\hat{m}(t) \times \hat{p} \times \hat{m}(t)$ where $\hat{m}(t)$ is the orientation of the FL moment. In a strongly non-collinear configuration such as in our STNO, it is helpful to resolve $\vec{\tau}_{st}(t)$ into a time-independent component $\hat{m}_0 \times \hat{p} \times \hat{m}_0$ (where $\hat{m}_0$ is the precession axis) and a time-varying component $\vec{\tau}_{st}(t) - \hat{m}_0 \times \hat{p} \times \hat{m}_0$; the time-independent part primarily determines the orientation of $\hat{m}_0$ while the time varying part contributes the anti-damping torque and can excite FL oscillations. For small angle precession, these components are simply the first and second terms in the Taylor series for $\vec{\tau}_{st}$ expanded about $\hat{m}_0$, respectively. Since for the total spin torque one has approximately $\tau_{st} \propto \sin\varphi_o$ where $\varphi_0$ is the offset angle between the FL and RL magnetizations, the strength of the anti-damping torque is approximately $\propto \cos\varphi_o$, and this results in $I_c = I_{c0} / |\cos\varphi_o|$ as employed in Fig. S2. Because increasing $H_y$ decreases $\varphi_0$, this angle-dependence of the anti-damping torque also explains why, as illustrated in the macrospin modeling results shown in Fig. S1c, increasing $H_y$ causes an increased $I_c$ and a decreased level of $p_0$ for a given value of $I/I_c$.

The macrospin simulations cannot however explain quantitatively the origin of the strong non-linear enhancement of $\Gamma_{eff}$, or equivalently the substantial non-linear reduction of ST efficiency ($\propto I \partial p / \partial I$) in the high $H_y$ regime as seen experimentally (compare Fig. S1c with Fig. 1d of the main text.) This is because the non-linear damping behavior has its origin in part in the significant spatial variation of the orientation of the magnetization of both the FL and the RL in this field bias regime, and to properly model that aspect of the dynamics requires micromagnetic simulations as discussed in S.7 below. In our macrospin simulations, there is decreased anti-damping when the precession grows large enough that the offset angle between the FL and RL approaches 90°, but as this occurs the precession axis also shifts toward this 90° angle, which has the effect of reducing the negative feedback such that $\Gamma_{eff}$ remains relatively small, with $\Gamma_{eff} / 2\pi < 0.2$ ns$^{-1}$. In the micromagnetic simulations, the spatial variation in the orientation of the magnetization, particularly the out-of-plane component of the magnetization at the two bottom ends of the elongated RL, allows a sharper onset for the reduced anti-damping without a large shift in the average precession axis (see S. 7).

## S.4. Tapered spin valve STNO fabrication process

The STNO was fabricated from a thin film multilayer stack of Py(5)/Cu(120)/Py(5)/Cu(12)/Py(20)/Cu(2)/Pt(30) (thicknesses in nm) deposited on an oxidized Si substrate, where Py = $Ni_{80}Fe_{20}$. The bottom Py(5) layer was to promote adhesion and played no significant magnetic or electrical role. The device had an elliptical cross-sectional area of ~ 50 × 145 $nm^2$ at the bottom of the nanopillar, as measured by scanning electron microscopy (± 5 nm). In this structure the thinner FL is located closer to the substrate than the thicker ferromagnetic reference layer (RL) and hence the latter has a higher aspect ratio than the former due to the sidewall tapering (20°-30°) during the ion-mill process. Therefore the shape anisotropy field ($H_k$) of the RL is much higher than for the FL, which fairly strongly fixes the unpinned RL, with a measured coercivity $H_c$(RL) ≈ 1300 Oe, greater than the $H_y$ range employed in the experiments, while $H_c$(FL) ≈170 Oe.

## S.5. Experimental estimation of the STNO's normalized oscillator power $p_0$ and nonlinear coupling $v$

Utilizing Eq. (1) of the main text, together with $E_0 = 2\pi V f^2 p_0 / \gamma^2$ as determined in section S.1, and employing the appropriate materials parameters for the Py free layer ($\alpha_G = 0.01$, $M_s = 560$ emu/cm$^3$) we have $\Delta f_{pred} \approx \left[6.86(1+v^2)/f^2 p_0\right]$ MHz at $T = 300$ K for our STNO, where $f$ is the oscillator frequency in GHz. To compare the measured linewidth $\Delta f_{meas}$ with $\Delta f_{pred}$ we determined values for $p_0$ under various field and current biases based on the measured power ($P_L$) delivered to the 50 Ω ($R_L$) transmission line, that is $P_L = \frac{1}{2}V_g^2 \frac{R_L}{(R_L+R_S)^2}$ where $R_S$ is the device resistance ($R_S \approx 25$ Ω) and $V_g$ is the amplitude of the generated microwave signal ($\Delta V(t) = V_g \sin(\omega t)$). A correction was also made for the calculated power lost to the Si substrate. We assumed that the magnetoresistance voltage signal is $\Delta V(t) = I \cdot R(t) \approx I \cdot (\partial R / \partial \varphi|_{\varphi_o}) \cdot \sin(\omega t) \cdot \varepsilon$ (as appropriate for a single-mode 1$^{st}$ harmonic and for the case $\varphi_o \leq 135°$) where $R(\varphi(t)) = (\Delta R_o) \left( \frac{\sin^2\left(\frac{\varphi(t)}{2}\right)}{1+\chi \cos^2\left(\frac{\varphi(t)}{2}\right)} \right)$ and $\varphi(t) = \varphi_o + \varepsilon \sin(\omega t)$. We measured $\Delta R_o = 0.2$ Ω and assumed $\chi = 1$ as previously reported [4] for Py.

We converted the measured $P_L$ to $p_0$ using the simplifying assumption that the FL moment is precessing in a parabolic potential and hence that $p_0 = \sin^2(\varepsilon/2)$. We find in the small power limit that $\varepsilon \approx \sqrt{\frac{2P_L}{R_L} \frac{R_L+R_S}{I \cdot R'(\varphi_o)}}$ with $R'(\varphi_o) = dR/d\varphi|_{\varphi_o}$. We determined the equilibrium offset angle $\varphi_0$ between the FL and the RL through the use of the predicted [8,9] $\varphi_0$ dependence of the onset current, $I_c = I_{c0} / |\cos\varphi_0|$ (see section S.3). In Fig. S2 we plot $I_c$ vs. $H_y$, where $I_c$ is determined by extrapolation of $p_N^{-1}$ to zero in the sub-threshold regime [10], along with the value of $\varphi_o$ for each $H_y$ as indicated by the fit to the predicted variation of $I_c$. As an example, we measure $P_L \approx 225$ pW at $H_y^{opt} = 700$ Oe and $I^{opt} = -4$ mA. From the measured onset current of ~ -2.5 mA for $H_y = 700$ Oe, we estimate $\varphi_o \approx 126°$ (see Fig. S2), and thus obtain $\varepsilon \approx$ 28.5°. We note however that both macrospin and micromagnetic modeling indicate that the orientation of the precession axis $\varphi_o$ shifts toward 90° as a function of increasing bias current (and increasing oscillation power) in this field regime due to the non-parabolic nature of the magnetic energy potential. (This is beneficial in providing some agility in the regime where the anisotropy fields are fairly closely balanced.) Thus for $I^{opt} = -4$ mA $\varphi_o$ could be as small as ~ 116° in this example. This would nevertheless result in only a negligible correction ($\varepsilon \approx 29.1°$). Similarly, while the determination of $I_c$ has a significant uncertainty due to the fact that we assumed macrospin behavior in estimating $I_c$ from the variation of $p$ with $I$ in the threshold region [10] while the MMS indicates a gradual onset even in the absence of thermal fluctuations (see Fig. 3a, Main Text) this results in only a small uncertainty on the determination of $\varphi_o$. For $\varepsilon \approx 29°$, we have $p_0 \approx 0.063$. At $I = -4.0$ mA we measure $f \approx 5.854$ GHz and $df/dI \approx -16$ MHz/mA, which yields $\nu \approx -0.65$ and thus $\Delta f_{pred} \approx 4.6 \pm 1.3$ MHz, quite close to $\Delta f_{meas} \approx 5 \pm 2$ MHz.

**S.6. Bias dependence of the power restoration rate $\Gamma_p$**

The power restoration rate $\Gamma_p \equiv p\Gamma_{eff}$ characterizes the dynamic damping of a STNO. As shown in Fig. S3, for the device discussed in the main text $\Gamma_p/2\pi$ ranges from 0.3 ns$^{-1}$ to 0.02 ns$^{-1}$ for 600 Oe < $H_y$ < 850 Oe. This rapid relaxation of power fluctuations is the result of the very strong $\Gamma_{eff}$ in our device configuration despite its relatively low power $p$. Above 900 Oe, $\Gamma_p$ increases rapidly despite the rapid decrease in $p$ due to the very strong enhancement in $\Gamma_{eff}$ ($dp/dI \to 0$) as $\varphi$ approaches 90°, indicating that here deviations in oscillator amplitude very quickly stabilize to the mean precession orbit, as previously implied by micromagnetic simulations [9]. We also plot the oscillator linewidth $\Delta f_{meas}$ in Fig. S3 for comparison to $\Gamma_p$. Note that above 850 Oe the two are comparable which explains why in this field regime $\Delta f_{meas}$, while increasing rapidly due to the decreasing $p$, becomes progressively less than predicted by the nonlinear theory [1,3] (see Fig. 2c) since the derivation of Eq. (1) requires $\Delta f \ll \Gamma_p/2\pi$.

## S.7. Micromagnetic Simulations of the STNO

To gain further understanding about the origins of the enhanced dynamic effective damping and the coherent oscillations in our STNO design, we performed zero temperature micromagnetic simulations [11] (MMS) of the idealized elliptical STNO. The micromagnetic simulations utilized the LLGS equation appropriate for a spin valve structure at $T = 0$ with the exchange constant $A = 13 \times 10^{-12}$ J/m, saturation magnetization $M_s = 560$ emu/cm$^3$, Gilbert damping parameter $\alpha = 0.01$, spin polarization [12] $P = 0.37$ and the volume discretized into 2.5 nm cubes for computational purposes. The simulations assumed a uniform (stepped) taper of 20 degrees, and an ideal elliptical shape with major and minor dimensions of 50 nm × 145 nm as indicated in Fig. 3b of the main text. Static ($I = 0$) simulations of a spin valve structure were used to determine the initial micromagnetic state of the FL and RL layers at the desired hard axis magnetic field. Dynamic ($I \neq 0$) simulations included effects from magnetic interactions between the two layers and the non-uniform circumferential Oersted field ($H_{Oe}$) generated by the bias current. ST was exerted upon both layers, with the local spin polarization of the projected current density incident upon a layer being dependent on the local magnetization vector of the other ferromagnet, *i.e.* the current flow was assumed to be quasi-one-dimensional [11,13,14]. We used the simplifying assumption that spins transmit the parallel component and reflect the antiparallel component of the local magnetization perfectly, depending on the direction the electrons traverse.

As discussed in the main text, a notable feature of the static magnetic configuration as calculated by the MMS in the field regime of interest $H_y = 700$ -1000 Oe is the substantial out-of-plane component of the magnetization $M_z$, at the bottom ends of the RL that arises from the taper, which according to the MMS can be as large as + 12% (-12%) of the total magnetization at the right (left) ends of the RL for the modeled structure (see Fig. 3b). Of course, the greater the taper the greater $|M_z|$ at the bottom ends. Due to this non-uniform out-of-plane magnetization (OPM) of the RL, when a current is applied such that the net flow direction of electrons is from the RL layer to the FL, there is significant spatial variation in the direction of polarization of the current incident on the FL, which we suggested in the main text is the origin of the enhanced non-linear dynamic damping of the ST induced oscillations of the FL, a suggestion that we expand on below.

In general, when current biases from -3 mA to -5 mA were applied in the MMS for hard axis biases $H_y$ = 700 -1000 Oe, we consistently obtained quite coherent oscillations of the simulated FL, when averaged over the FL and over many cycles, as illustrated for example by Mov. S1 for $I$ = -2.25 mA, Mov. S2 $I$ = -2.5 mA, Mov. S3 for $I$= -3.0 mA, Mov. S4 for $I$ =-4.0 mA, all with $H_y$= 800 Oe applied in the MMS. However, while the averaged MMS results are quite coherent, the micromagnetic details of the oscillations are significantly more complex than the uniform, in-plane "clam-shell" elliptical precession predicted by rigid domain, macrospin modeling [7]. The most notable aspect is that the magnetization does not oscillate in phase across the FL, particularly with respect to the $M_z$ component (see Fig. S4 and Movies). This effect apparently originates with the OPM at the ends of the RL in the initial static configuration, which induces a similar OPM in the ends of the FL, but one that is weaker due to the stronger out-of-plane demagnetization field in the thinner layer. When bias current is applied the spin torque exerted by the OP polarized component of the current acts to enhance $M_z$ in the FL's two end regions, with the time averaged $\bar{M}_z$ = 0 only in the (diagonal) center region of the FL. This reduces the overall efficiency of the in-plane ST in exciting the in-plane precessional mode, raising $I_c$ in comparison to that predicted for a rigid domain, and, once $I > I_c$, resulting in the in-plane precession in the right end region leading (lagging) that in the left end during the half cycle for which the time-varying demagnetization field in that end is enhanced (reduced) by the effect of the OP ST. Since the OP ST and IP ST increase together with $I$, the result at first is a slower than otherwise expected rate of increase of $p_0$ with bias, and hence a higher effective dynamic damping of the oscillation (see Fig. S4a and Mov. S1 for $I$ = -2.25 mA and Fig. 3b).

Eventually, at high enough bias, *e.g.* $I \approx$ - 2.5 mA in the simulated device, the in-plane precession amplitude becomes large enough that the exchange coupling across the FL results in $\text{sgn}(M_z)$ being the same across the FL for much of two half cycles (see Fig. S4b and Mov. S2). We suggest that this enhances the efficiency of the IP ST and is the origin of the increased $dp/dI$, which reduces the non-linear dynamic damping in this bias region, $\Gamma_{\textit{eff}}/2\pi \approx 0.4$ GHz (see Fig. 3a), although still keeping it higher than expected from macrospin modeling, $\Gamma_{\textit{eff}}/2\pi \approx \Gamma_0/2\pi \approx 0.1 - 0.2$ GHz.

As the bias is increased further and the amplitude of the oscillation grows, the fraction of the oscillation's half period when $\text{sgn}(M_z)$ is not uniform across the FL becomes quite small, at about the point where the extremum of the oscillation in the hard axis direction, $\varphi_{\min}$, brings the relative local orientation between the FL and the RL to $\varphi_{\min} \approx 90^o$. In the MMS this occurs for $I \approx$ - 3.0 mA (see Mov. S3). At that precession amplitude, for which $\varepsilon \approx 30^o$, the effectiveness of the IP anti-damping ST on the FL decreases rapidly with further increases in $\varepsilon$ while the positive damping $\Gamma_+$ continues to increase, which reduces the differential ST effect of increasing $I$. For still higher currents the MMS shows that $\varphi < 90^o$ for a significant part of the oscillation period. During that time the electrons with polarization anti-parallel to the local orientation of the FL magnetization that are reflected back to the RL exert an anti-damping ST on the two RL ends which, the MMS indicates, excites them into substantial oscillation about their local effective fields that point partially out of plane. This oscillation of the RL ends, which is not in phase with the main FL oscillation, results in time varying changes in the strength of the OP ST exerted on the FL and hence in a substantial increase in the variation of $\text{sgn}(M_z)$ across the FL during the precession (see Fig. S4c and Mov. S4). The overall outcome in the MMS is a major increase in nonlinear effective damping (see Fig. 3a). The ST induced oscillations of the RL ends grow in amplitude with increased bias and the coherency of the FL in-plane oscillation gradually degrades, which we suggest is the origin of the broadening of the linewidth in the experimental device illustrated in Fig. 2a of the main text for the case of $H_y$ =700 Oe, and $|I|$ > 4.0 mA. When MMS are performed with higher values of $H_y$, higher levels of $I$ are required, both to initiate steady state oscillation and to bring $\varphi_{\min} < 90°$ and hence cause quasi-saturation in $p_0$, in general accord with experiment.

      Our study of the details of the MMS for this tapered, and non-collinear spin valve structure indicates that one effective method for obtaining a high non-linear effective damping and hence quasi-linear behavior in a STNO is to apply a non-uniform spin-polarized current to the oscillating FL where the current locally has a significant polarization component that is orthogonal to the precessional mode. This, we suggest, has the effect of introducing phase gradients in the dominant oscillator mode, which is in-plane precession in this case. This establishes an enhanced non-linear damping without immediately resulting in the generation of a second, strongly competing mode that would result in a unacceptably low phase stability (broad linewidths.) We suggest that a similar approach can be employed when the ST precessional mode is of a different character; the essential point, we argue, is to utilize a non-uniform ST to

both excite the precessional mode and to dampen the variation of its amplitude with bias, that is to enhance the nonlinear damping, within the desired operating regime.

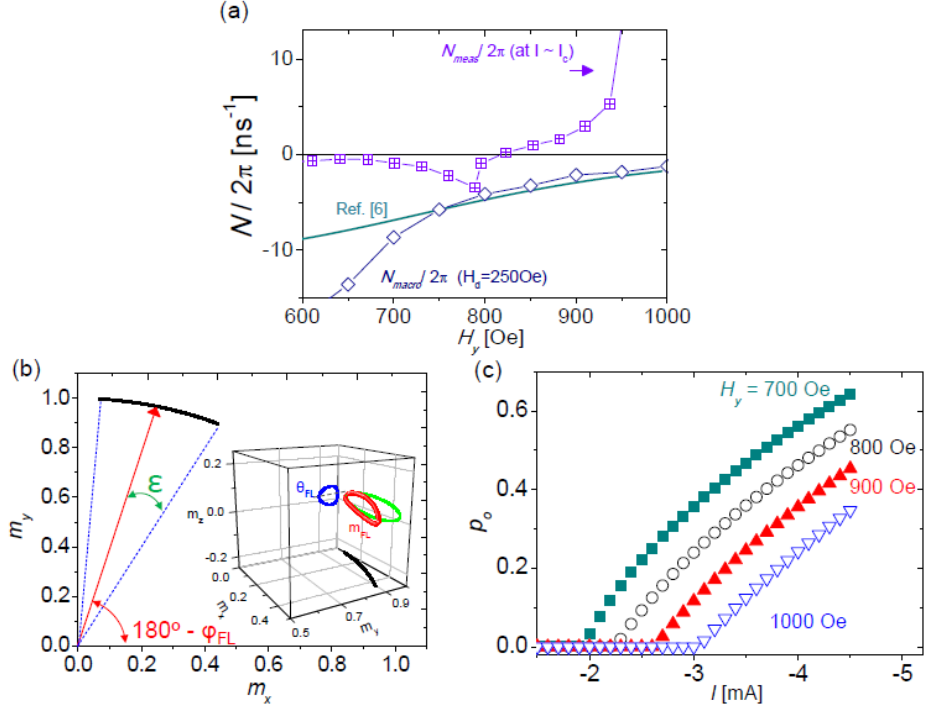

Fig. S1. (a) Result of a macrospin simulation for the agility $N/2\pi$ as a function of $H_y$ as determined just above the onset of oscillation, $p_0 \approx 0.01$, for the case where $H_k = 450$ Oe, $H_d = 250$ Oe and where the incident spin polarization $\hat{p} = (-1,0,0)$. Also shown is the agility prediction from the analysis of Ref. 6, for the same applied and anisotropy fields, and the experimental result for the STNO discussed in the main text as measured just above the critical current for the onset of oscillation as a function of $H_y$. (b) Example of the in-plane ($m_x$, $m_y$) trajectory and the 3-dimensional ($m_x$, $m_y$, $m_z$) trajectory (inset) of the FL moment for $H_y = 800$ Oe, $H_d = 250$ Oe, $H_k = 450$ Oe, $I = -2.3$ mA. (c) The normalized power ($p_o = \sin^2(\varepsilon/2)$) as a function of $I$ and $H_y$ for $H_d = 250$ Oe, $H_k = 450$ Oe.

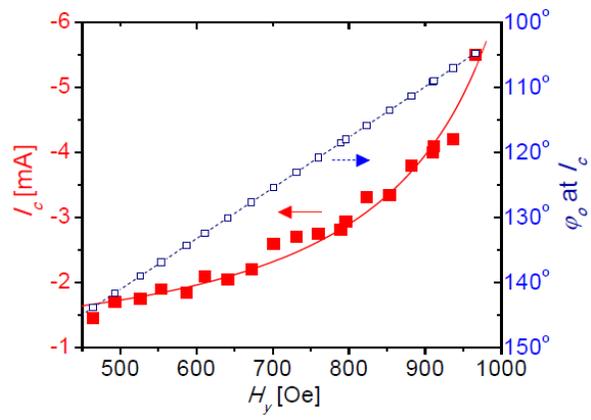

Fig. S2. $I_c$ for the onset of the auto-oscillation and the value of the initial offset angle $\varphi_o$ determined from $I_c$, both as a function of $H_y$.

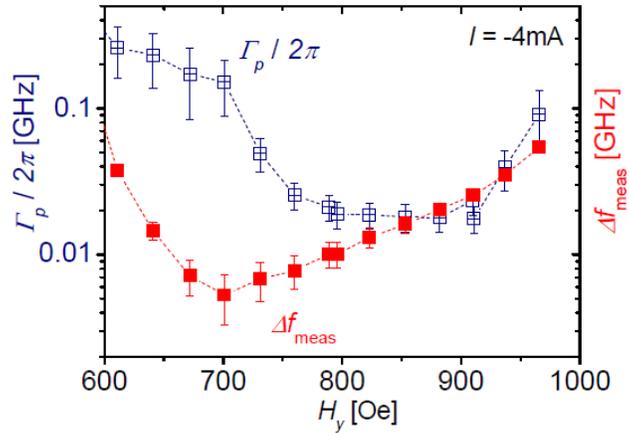

Fig. S3. Dynamic damping parameter $\Gamma_p/2\pi$ $(\equiv p_0\Gamma_{eff}/2\pi)$ of the STNO as determined from the experimental data using Eq. S15 as a function of $H_y$, for $I = -4$ mA. Also shown are $\Delta f_{meas}$ for comparison.

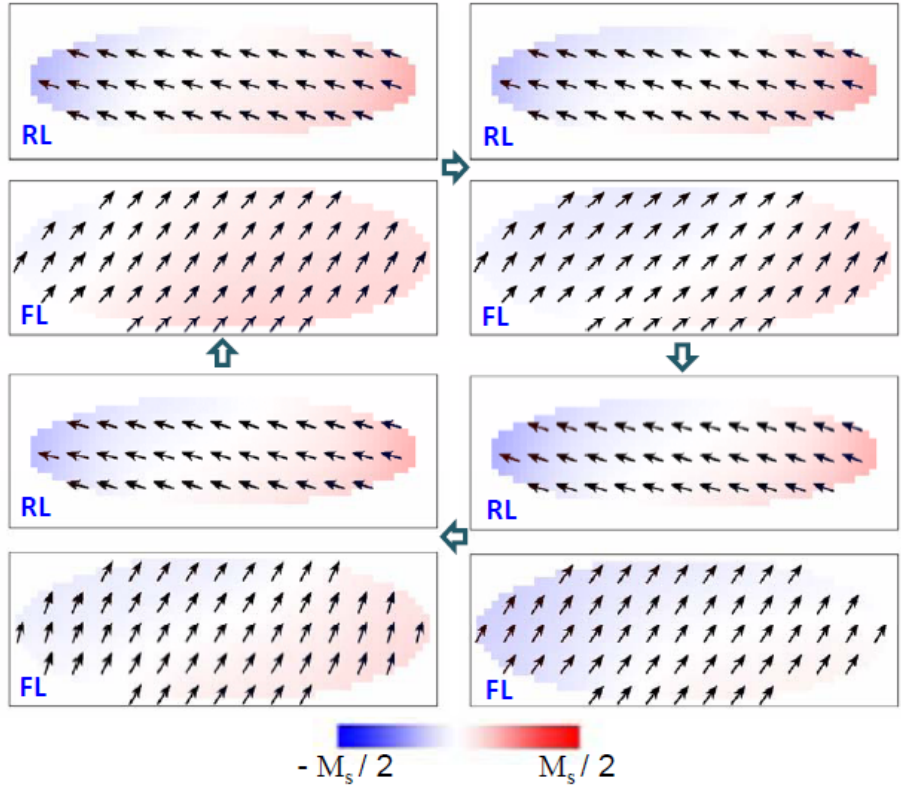

Fig. S4 (a) Time varying snapshots of the FL and RL magnetization during one cycle of the micromagnetic simulated oscillation for $H_y = 800$ Oe, $I = -2.25$ mA (see also Mov. S1).

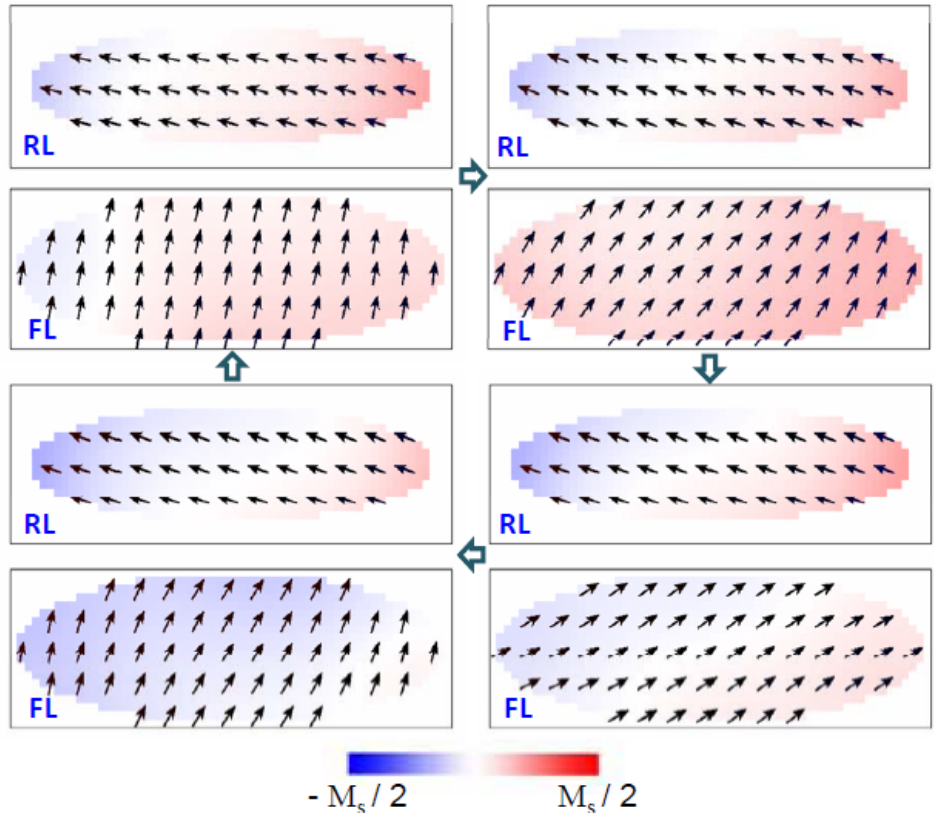

Fig. S4 (b) Time varying snapshots of the FL and RL magnetization during one cycle of the micromagnetic simulated oscillation for $H_y = 800$ Oe, $I = -2.5$ mA (see also Mov. S2).

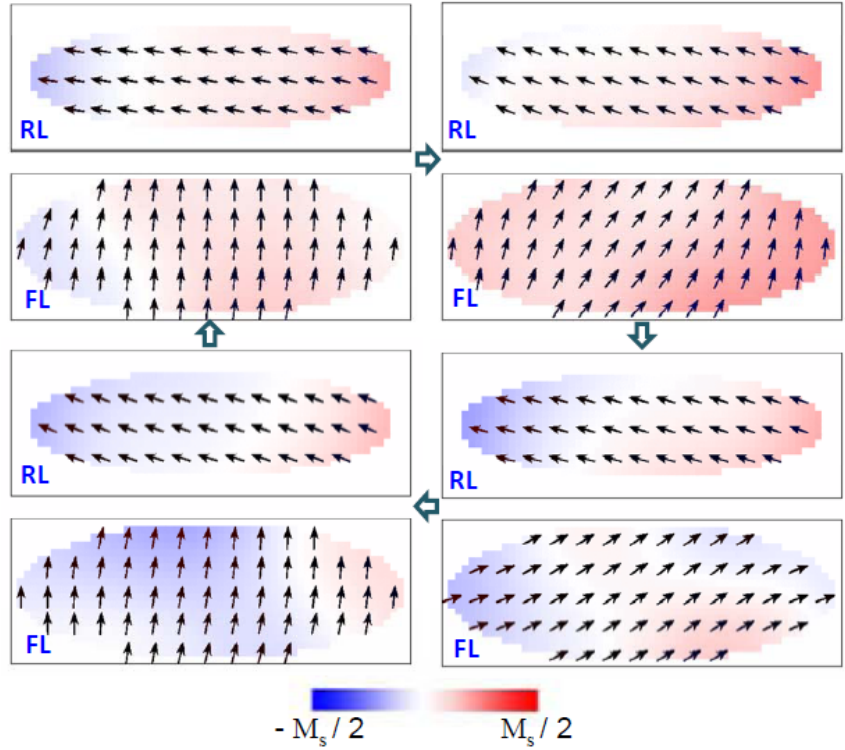

Fig. S4 (c) Time varying snapshots of the FL and RL magnetization during one cycle of the micromagnetic simulated oscillation for $H_y$ = 800 Oe, $I$ = -4.0 mA (see also Mov. S4). The ST applied to the center area drives the oscillations while the ST at the edges provides enhanced dynamic damping when the local orientation between the FL and RL magnetization is near 90°.

Mov.S1 Simulated micromagnetic oscillations for $H_y$ = 800 Oe, $I$ = -2.25 mA with $H_{oe}$. The top, smaller ellipse represents the magnetization of bottom interface (facing to the FL) of the RL layer while the bottom ellipse represents the magnetization at the top interface of the FL. (Time snapshots in Fig. S3a).

Mov. S2 Simulated micromagnetic oscillations for $H_y$ = 800 Oe, and a higher bias current $I$ = -2.5 mA with $H_{oe}$. (Time snapshots in Fig. S3b).

Mov. S3 Simulated micromagnetic oscillations for $H_y$ = 800 Oe, and a higher bias current $I$ = -3.0 mA with $H_{oe}$.

Mov. S4 Simulated micromagnetic oscillations for $H_y$ = 800 Oe, and a higher bias current $I$ = -4.0 mA with $H_{oe}$. (Time snapshots in Fig. S3c).

## Supplemental References